\documentclass[fleqn,usenatbib]{mnras}

\usepackage{newtxtext,newtxmath}

\usepackage[T1]{fontenc}

\DeclareRobustCommand{\VAN}[3]{#2}
\let\VANthebibliography\thebibliography
\def\thebibliography{\DeclareRobustCommand{\VAN}[3]{##3}\VANthebibliography}

\usepackage{graphicx}	
\usepackage{amsmath}	
\usepackage{url}
\usepackage{enumitem}

\newcommand{\kabs}{\kappa_{\text{abs}}}
\newcommand{\ksca}{\kappa_{\text{sca}}}
\newcommand{\kext}{\kappa_{\text{ext}}}

\newcommand{\matr}[1]{\mathbf{#1}}

\title[Scattering of Large Irregular Grains]{(Sub)millimeter Dust Polarization of Protoplanetary Disks from Scattering by Large Millimeter-Sized Irregular Grains}

\author[Z.-Y. D. Lin et al.]{
Zhe-Yu Daniel Lin,$^{1}$\thanks{Jefferson Fellow. E-mail: zdl3gk@virginia.edu}
Zhi-Yun Li,$^{1}$
Haifeng Yang,$^{2}$
Olga Mu\~noz,$^{3}$
Leslie Looney,$^{4}$
Ian Stephens,$^{5}$ 
\newauthor
Charles L. H. Hull,$^{6,7}$\thanks{NAOJ Fellow}
Manuel Fern\'andez-L\'opez,$^{8}$
and Rachel Harrison$^{4}$
\\
$^{1}$Department of Astronomy, University of Virginia, 530 McCormick Rd., Charlottesville, Virginia 22904, USA\\
$^{2}$Kavli Institute for Astronomy and Astrophysics, Peking University, Yi He Yuan Lu 5, Haidian Qu, Beijing 100871, People's Republic of China \\
$^{3}$Instituto de Astrof\'isica de Andaluc\'ia, CSIC Glorieta de la Astronom\'ia s/n, E-18008 Granada, Spain \\
$^{4}$Department of Astronomy, University of Illinois, 1002 W Green St., Urbana, IL 61801, USA \\
$^{5}$Department of Earth, Environment and Physics, Worcester State University, Worcester, MA 01602, USA \\
$^{6}$National Astronomical Observatory of Japan, NAOJ Chile Observatory, Alonso de C\'{o}rdova 3788, Vitacura, Santiago, Chile \\
$^{7}$Joint ALMA Observatory, Alonso de C\'ordova 3107, Vitacura, Santiago, Chile \\
$^{8}$Instituto Argentino de Radioastronom\'ia (CCT-La Plata, CONICET; CICPBA), C.C. No. 5, 1894, Villa Elisa, Buenos Aires, Argentina
}

\date{Accepted XXX. Received YYY; in original form ZZZ}

\pubyear{2015}

\begin{document}
\label{firstpage}
\pagerange{\pageref{firstpage}--\pageref{lastpage}}
\maketitle

\begin{abstract}
%

The size of dust grains, $a$, is key to the physical and chemical processes in circumstellar disks, but observational constraints of grain size remain challenging. (Sub)millimeter continuum observations often show a percent-level polarization parallel to the disk minor axis, which is generally attributed to scattering by $\sim 100\mu$m-sized spherical grains (with a size parameter $x \equiv 2\pi a / \lambda < 1$, where $\lambda$ is the wavelength). Larger spherical grains (with $x$ greater than unity) would produce opposite polarization direction. However, the inferred size is in tension with the opacity index $\beta$ that points to larger mm/cm-sized grains. We investigate the scattering-produced polarization by large irregular grains with a range of $x$ greater than unity with optical properties obtained from laboratory experiments. Using the radiation transfer code, RADMC-3D, we find that large irregular grains still produce polarization parallel to the disk minor axis. If the original forsterite refractive index in the optical is adopted, then all samples can produce the typically observed level of polarization. Accounting for the more commonly adopted refractive index using the DSHARP dust model, only grains with $x$ of several (corresponding to $\sim$mm-sized grains) can reach the same polarization level. Our results suggest that grains in disks can have sizes in the millimeter regime, which may alleviate the tension between the grain sizes inferred from scattering and other means. Additionally, if large irregular grains are not settled to the midplane, their strong forward scattering can produce asymmetries between the near and far side of an inclined disk, which can be used to infer their presence.

\end{abstract}

\begin{keywords}
polarization -- protoplanetary disks -- circumstellar matter
\end{keywords}



\section{Introduction}

Dust in circumstellar disks only holds about 1$\%$ of the total disk mass, yet it plays a key role in shaping disk properties and serves as the fundamental building blocks of planets. One of the most relevant properties of dust is its size. The growth to planets directly requires the aggregation of grains from submicron sizes inherited from the interstellar medium to kilometer-sized planetesimals and eventually to planets \citep[e.g.][]{Bitsch2015, Drazkowska2018}. The grains dominate the opacity, which is sensitive to grain size \citep[e.g.,][]{Draine2006, Birnstiel2018}. As a result, grains and how the various sizes distribute in the disk affect the temperature structure \citep[e.g.][]{DAlessio2001, Inoue2009, Williams2011} and the multiwavelength observational appearance of disks \citep[e.g.][]{Dong2018, Huang2020, Sierra2021}. The grain sizes directly impact the chemistry of disks because
of its dependence on temperature \citep[e.g.][]{Gavino2021} and dust opacity at UV which affects photodissociation \citep[e.g.][]{Cleeves2011}, and also because surface chemistry relies on the grain surface area \citep[e.g][]{Harada2017}. The dynamics and evolution of disks depend on the grain sizes which govern how coupled the grains are to the gas and also the level of ionization \citep[e.g.][]{Hu2021}. Despite the importance of dust grain size, it has been difficult to directly constrain their sizes from observations.

One way to measure the grain size of disks is through (sub)millimeter continuum polarization. With the tremendous sensitivity of the Atacama Large Millimeter/submillimeter Array (ALMA), (sub)millimeter polarized images have been resolved for many disks. Grains scatter radiation, and the scattered radiation becomes polarized. Since the grains scatter its own thermal radiation at the (sub)millimeter wavelengths, it is often called self-scattering \citep{Kataoka2015}. A characteristic feature of polarization due to self-scattering is the unidirectional polarization that is parallel to the minor axis of inclined disks \citep{Yang2016, Kataoka2016}. The majority of the observed polarization images are $\sim 1\%$ polarized and show the unidirectional polarization feature, especially at relative short wavelength ALMA Bands, like Bands 6 and 7 \citep[e.g.][]{Stephens2017, Bacciotti2018, Hull2018, Lee2018, Dent2019, Mori2019, Sadavoy2019, Stephens2020, Aso2021, Harrison2021}. To efficiently produce polarization, the grain size, $a$, must not be much smaller than the observing wavelength, $\lambda$, \citep{Kataoka2015}, but grains much larger than the wavelength can cause the polarization to become parallel to the disk major axis \citep{Yang2016}. In other words, the size parameter, $x \equiv 2 \pi a / \lambda$, should be of order unity. As a result, polarization is deemed sensitive to grain size and the majority of detected polarization images have been taken as evidence for grains of $\sim 100\mu$m \citep[e.g.][]{Yang2016, Kataoka2017, Ohashi2018, Ohashi2020, Lin2020, Ueda2021}.

Another way to estimate grain sizes is through the wavelength dependence of opacity. For grains with absorption opacity that goes as $\kabs \propto \nu^{\beta}$, the opacity index $\beta$ depends on the grain size \citep{Draine2006}. For the interstellar medium, the small grains ($\sim 0.1 \mu$m) have $\beta \sim 1.7$ \citep{Weingartner2001}. The typically inferred $\beta$ of disks at millimeter wavelengths is $\sim 1$ or lower \citep[e.g.][]{Beckwith1990, Ubach2012, Carrasco2016, Sheehan2018, Macias2019, Tobin2020, Lin2021}. At face value, the low $\beta$ suggests mm/cm-sized grains \citep[e.g.][]{Draine2006, Testi2014} which directly contradicts the size inferred from polarization. However, recently it has been pointed out that the correspondence between $\beta$ and the grain sizes is actually much more complicated when one takes into account the effects of the scattering whose contribution to the dust opacity is not negligible in the case of disks \citep[e.g.][]{Zhu2019, Sierra2020}. More detailed studies of the millimeter spectrum of disks including scattering opened the possibility for $\sim$ hundred-$\mu$m grains \citep[e.g.][]{Huang2020, Liu2019, Lin2020, Ueda2020, Ueda2021, Ueda2022, Sierra2021}, but many still resulted in grains in the mm/cm-sized regime \citep[e.g.][]{Carrasco2019, Ohashi2020, Liu2021, Macias2021, Sierra2021}.

The orders of magnitude discrepancy between $100$ $\mu$m versus the mm/cm regime can heavily affect the interpretation of the physical and chemical properties of the disk given the importance of the grain size. In this paper, we demonstrate that the perceived accuracy of polarization measurements in constraining the grain size is in part due to the strict assumption of spherical grains. Since grains coagulate to form larger grains, the shape of grains is expected to be irregular \citep[e.g.][]{Krause2004, Ormel2007, Blum2008}. Though the assumption of spherical grains is largely based on its numerical simplicity calculated from Mie theory \citep{Mie1908}, the scattering properties of spherical grains becomes drastically different from those of irregular grains once the size becomes comparable to the size of the observing wavelength, as predicted from more sophisticated numerical techniques \citep[e.g.][]{Shen2008, Shen2009, Kirchschlager2013, Kirchschlager2014, Tazaki2016, Tazaki2018, Tazaki2019, Kirchschlager2020} and shown from experimental measurements \citep[e.g.][]{Munoz2011, Munoz2021}. 

In this paper, we use scattering matrices measured from the Instituto de Astrof\'isica de Andaluc\'ia (IAA) Cosmic Dust Laboratory \citep[CoDuLab;][]{Munoz2011, Munoz2012, Munoz2021} as illustrative samples of the scattering matrix when the size parameter $x$ is much larger than unity to simulate the (sub)millimeter disk polarization. By using the experimentally measured scattering matrices, we can consider grains with size parameters up to $575$ in this paper, which is larger than what current numerical techniques can readily achieve. In Section~\ref{sec:setup}, we briefly describe the properties of the experimental dust samples and disk model setup. We use the Monte Carlo radiative transfer code RADMC-3D\footnote{RADMC-3D is available at \url{https://www.ita.uni-heidelberg.de/~dullemond/software/radmc-3d/}} to simulate the polarization images at millimeter wavelengths \citep{Dullemond2012}. Section~\ref{sec:results} presents the simulated polarization images comparing the use of laboratory measured scattering matrix and the Mie calculations. We show that irregular grains with large size parameters can still produce polarization parallel to the disk minor axis. Since large grains exhibit strong forward scattering, we also show how forward scattering affects the polarization image. Section~\ref{sec:discussion} offers a discussion of the implications and the results are summarized in Section~\ref{sec:conclusion}. 

\section{Simulation Setup} \label{sec:setup}

\subsection{Dust Model} \label{sec:dustmodel}
We use the experimental scattering matrix for a set of forsterite (in the form of (Mg, Fe)$_{2}$ SiO$_{4}$ and Mg$_{3}$Si$_{2}$O$_{5}$(OH)$_{4}$) samples presented by \cite{Munoz2021}. The elements of the scattering matrix $F_{ij}$ depend on the physical properties of the grain, such as shape, size and composition, and the direction of scattering (e.g., the angle form by the directions of the incident and scattered beams). In the case of randomly oriented particles as is the case in the CoDuLab experiment, all scattering planes are equivalent and the scattering direction is fully described by the scattering angle $\theta$. The $4 \times 4$ scattering matrix $\matr{F}$ becomes a block diagonal and is defined by 
\begin{align}
    \begin{pmatrix}
        I_{s} \\
        Q_{s} \\
        U_{s} \\
        V_{s}
    \end{pmatrix}
    \propto
    \begin{pmatrix}
        F_{11} & F_{12} & 0 & 0 \\
        F_{12} & F_{22} & 0 & 0 \\
        0 & 0 & F_{33} & F_{34} \\
        0 & 0 & - F_{34} & F_{44}
    \end{pmatrix}
    \begin{pmatrix}
        I_{i} \\
        Q_{i} \\
        U_{i} \\
        V_{i}
    \end{pmatrix}
\end{align}
where $(I_{i}, Q_{i}, U_{i}, V_{i})$ and $(I_{s}, Q_{s}, U_{s}, V_{s})$ are the Stokes parameters of the incoming and scattered light respectively. For unpolarized incident light ($(I_{i}$,$Q_{i}$,$U_{i}$,$V_{i}$)=$(1,0,0,0)$), the $F_{11}(\theta)$ function is proportional to the flux of the scattered light and is called the phase function. Note that the scattering opacity, $\ksca$, for randomly oriented particles is directly related to the $F_{11}$ element by
\begin{align} \label{eq:ksca_from_F11}
    \ksca \equiv 2 \pi \int_{0}^{\pi} F_{11}(\theta) \sin \theta d \theta.
\end{align}
Also, for unpolarized incident light, the ratio $-F_{12}(\theta) / F_{11}(\theta)$ is called the degree of linear polarization of the scattered light, hereafter DLP. 


Due to the limited amount of grain samples, the measurements are limited to the $F_{11}(\theta)$, $F_{12}(\theta)$, and $F_{22}(\theta)$ scattering matrix elements, whereas $F_{33}(\theta)$, $F_{34}(\theta)$, and $F_{44}(\theta)$ are not measured. We supplement the missing scattering elements by the following. Motivated by laboratory measured $F_{33}$ elements of irregular grains of olivine from \cite{Munoz2000} (see their Fig.~5), we set $F_{33}(\theta) = (-0.45 \theta + 1) F_{11}(\theta)$ with $\theta$ in radians. We do not consider circular polarization and set $F_{34}$ and $F_{44}$ to zero. Given that the ratio $F_{34} / F_{11}$ of irregular grains from \cite{Munoz2000} is $\sim 10\%$ at most, circular polarization is at least an order of magnitude less than linear polarization which makes the impact marginal. Since there is a lack of confident detection in Stokes $V$ in disks \citep[e.g.][]{Stephens2017}, it is beyond the scope of this paper. 


The bulk forsterite sample was processed for producing various size distributions, namely XS, S, L and XL. As described in \cite{Munoz2021}, sample XS represents grains with sizes in the transition region between the Rayleigh and resonance scattering regimes, which has $x$ of several. Samples S and L belong to the resonance and/or transition region between the resonance and geometric optics regimes, which have $x$ $\sim 20$ and $\sim 40$. And sample XL consists of particles with $x\sim 600$ which represents the geometric optics regime. Table~\ref{tab:sample} lists the effective size parameter ($x_{\text{eff}}$) of each sample at the experimental wavelength ($\lambda=514$~nm) which was used to measure the scattering properties in the laboratory. The corresponding equivalent effective radii $a_{\text{eff}}$ at $1$~mm wavelength, which is the wavelength we use for the simulations, are also presented by fixing the effective size parameter through $a_{\text{eff}} = x_{\text{eff}} \times 1 \text{mm} / ( 2 \pi )$. At $1$~mm wavelength, the samples XS and S are representative of mm-sized grains, while the samples L and XL are in the centimeter and decimeter size regime.

Since the experimental scattering matrices do not cover the full angular extent ($3^{\circ}$ to $177^{\circ}$), we use the synthetic scattering matrix whose scattering angle $\theta$ is defined from $0^{\circ}$ to $180^{\circ}$ to adequately apply the results from experimental scattering matrix to radiation transfer. The extrapolation of the scattering matrix elements is based on the technique described by \cite{EscobarCerezo2017} and further improved by \cite{GomezMartin2021}.

To compare the disk images produced from experimental scattering matrix and from Mie calculations, we need the absorption opacity $\kabs$ and the absolute values for each element of the scattering matrix (including $\ksca$). However, the experimental data does not have $\kabs$ and its scattering matrix is a relative quantity, i.e., the absolute values of each element are unknown. Thus, for each experimental sample, we assign the same $\kabs$ and scattering opacity $\ksca$ from Mie theory.

Mie theory requires only two inputs\footnote{The Python code for Mie scattering is written by C. Dullemond which is included in the RADMC-3D package. The code is based on the FORTRAN code available at \url{https://www.astro.princeton.edu/~draine/scattering.html} which originates from \cite{Bohren1983}. }, the size parameter and refractive index, to produce the optical properties like $\kabs$ and $\ksca$ and its own scattering matrix (note we obtain the absorption and scattering efficiencies $Q_{\text{abs}}$ and $Q_{\text{sca}}$ which is related to the opacities in cross section per gram of dust through $\kabs=Q_{\text{abs}} \pi a^{2} / m_{g}$ and $\ksca = Q_{\text{sca}} \pi a^{2} / m_{g}$ where $m_{g}$ is the mass of the grain). Since each experimental sample has a size distribution, as opposed to a single size, we use the same size distribution for each respective sample which was derived from the original experiment \citep{Munoz2021}. To produce smooth profiles from Mie calculations, we refine the grain size bins by linearly interpolating the experimental size distributions; otherwise, the coarse grain size bins measured from the laboratory produces severe oscillations when implementing Mie calculations.

For the refractive index $m = n + i k$ where $n$ and $k$ are the real and imaginary parts respectively, we use $m = 1.65 + 10^{-5} i$ for the forsterite material at the experiment wavelength (at $\lambda=514$~nm; \citealt{Huffman1973}) to directly compare with the experimental scattering matrix. However, we note that the experimental $k$ can be a few orders of magnitude lower than the $k$ at millimeter wavelengths from commonly adopted material for disks, such as water-ice, silicates, or organics \citep[e.g.][]{Pollack1994, Draine2003, Birnstiel2018}, which we discuss in Section~\ref{sec:refractive_index}.

With the size parameter distribution and $m$ known, we can obtain $\kabs$ and $\ksca$ from Mie theory. We then scale $F_{11}$ of the experimental data such that $\ksca$ is equal to their Mie counterpart. Explicitly, this means 
\begin{align}
    F_{11}^{\text{scaled}} \equiv \dfrac{ \ksca^{\text{Mie}} }{ \ksca^{\text{synthetic}} } F_{11}^{\text{synthetic}} 
\end{align}
where $F_{11}^{\text{synthetic}}$ is the synthetic phase function derived from experimental data, $\ksca^{\text{Mie}}$ is the scattering opacity from Mie theory, $\ksca^{\text{synthetic}}$ is the scattering opacity from the synthetic phase function, and $F_{11}^{\text{scaled}}$ is the scaled phase function with the same scattering opacity as the Mie counterpart. The other scattering elements are scaled by the same factor.

In principle, the $F_{ij}^{\text{scaled}}$ can be used for the radiative transfer calculations. However, while conducting the Monte Carlo radiative transfer simulation, most photons are scattered in the forward direction given the large forward scattering peak of the large grains for the experimental samples (and also the Mie calculations). This leads to difficulty in obtaining smooth images since most of the radiation directed in the forward direction is not polarized and only the small portion of photons that are side-scattered contribute polarization. In other words, most of the (finite number of) photon packets in the simulation travel in the forward scattering region leaving only a portion of photon packets ($\sim$ a few percent; see the Appendix~\ref{sec:compare_truncation}) to contribute to the polarization image. Given that large forward scattering is effectively no scattering at all, we truncate the forward scattering peak at $1^{\circ}$ by setting 
\begin{equation} \label{eq:Fij_truncation}
    F_{ij}^{\text{truncated}}(\theta) = 
    \begin{cases}
        0 & \text{for } \theta < 1^{\circ} \\
        F_{ij}^{\text{scaled}}(\theta) & \text{for } \theta \geq 1^{\circ}
    \end{cases}
\end{equation}
(see, e.g., \citealt{Nakajima1988, Iwabuchi2006}). 
The truncation means the photon packets are forced to trace the scattering angles outside of $1^{\circ}$. This helps to obtain smooth polarization images with achievable computational time, since photon packets are no longer wasted in tracking the forward scattering direction. Note that $\ksca$ is recalculated based on the truncated $F_{11}$ since we are treating forward scattering as not having interacted with the medium at all (e.g., Eq.~\ref{eq:ksca_from_F11} always holds). We have experimented with different choices of the truncation angles and found quantitatively similar results (see Appendix~\ref{sec:compare_truncation}). The scattering elements of the results from Mie calculations are truncated as well. Hereafter, unless explicitly stated otherwise, the $\ksca$ and $F_{ij}$ of the experimental data were scaled and truncated and those from Mie calculations were truncated.

For convenience, the extinction opacity $\kext$ is defined as $\kext \equiv \kabs + \ksca$ and the albedo is $w \equiv \ksca / (\kabs + \ksca)$. In Section~\ref{ssec:disk_model_setup} below, we define the surface density in terms of the optical depth and as such $\kext$ cancels out. This means the absolute difference in opacity across the different samples would not matter since it is the optical depth (surface density multiplied by $\kext$) that determines the radiation transfer results.

\begin{figure}
    \centering
    \includegraphics[width=\columnwidth]{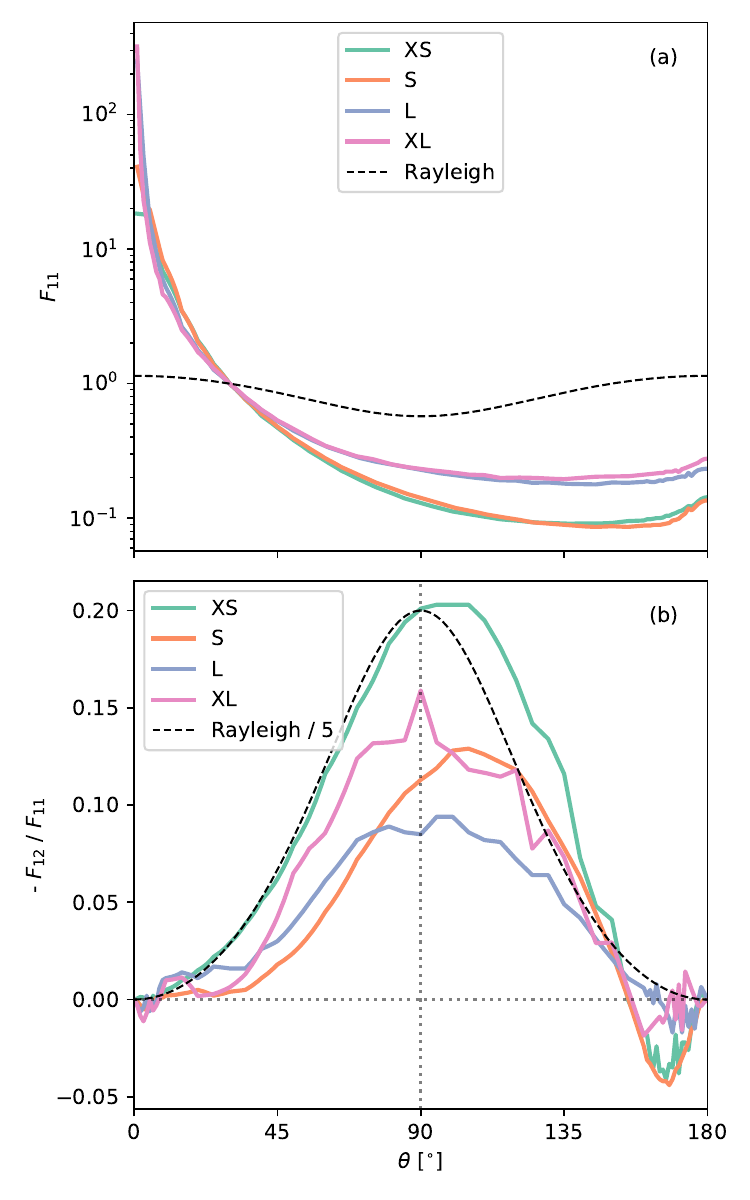}
    \caption{ 
        Top: The phase function $F_{11}$ as a function of scattering angle $\theta$ for the different samples. The Rayleigh scattering $F_{11} \propto \cos^{2}\theta + 1$ is also shown as a comparison. For better comparison, each are normalized at $\theta=30^{\circ}$ to an arbitrary value of 1. Bottom: The degree of linear polarization (DLP) which is defined as $- F_{12}/F_{11}$ for the different samples. The Rayleigh scattering DLP at $\theta=90^{\circ}$ should be $100\%$, but we scaled it down to $20\%$ for better comparison with the experimental data. 
    }
    \label{fig:samples_z11_dlp}
\end{figure}

Fig.~\ref{fig:samples_z11_dlp}a shows the experimental phase function $F_{11}$ as a function of the scattering angle $\theta$ for each of the samples. The element represents the angular distribution of scattered photons. For comparison, we also show $F_{11}$ for Rayleigh scattering which goes as $F_{11} \propto \cos^{2}\theta + 1$. The experimental $F_{11}$ for each sample shows a strong forward scattering which is different from the Rayleigh limit that has equal forward and backward scattering.

Fig.~\ref{fig:samples_z11_dlp}b shows the experimental $-F_{12} / F_{11}$ (i.e., the DLP). Positive values of DLP mean the polarization is perpendicular to the scattering plane. Across all four samples, the DLP curves are roughly bell-shaped with peaks of $\sim 0.1-0.2$. The samples XS and S have small but negative polarization for large $\theta$ (in the vicinity of back-scattering). The DLP for all four samples are similar to the bell-shaped curve for particles in the Rayleigh limit, except the peak in the Rayleigh limit is $100\%$ polarized. 

\begin{figure*}
    \centering
    \includegraphics[width=\textwidth]{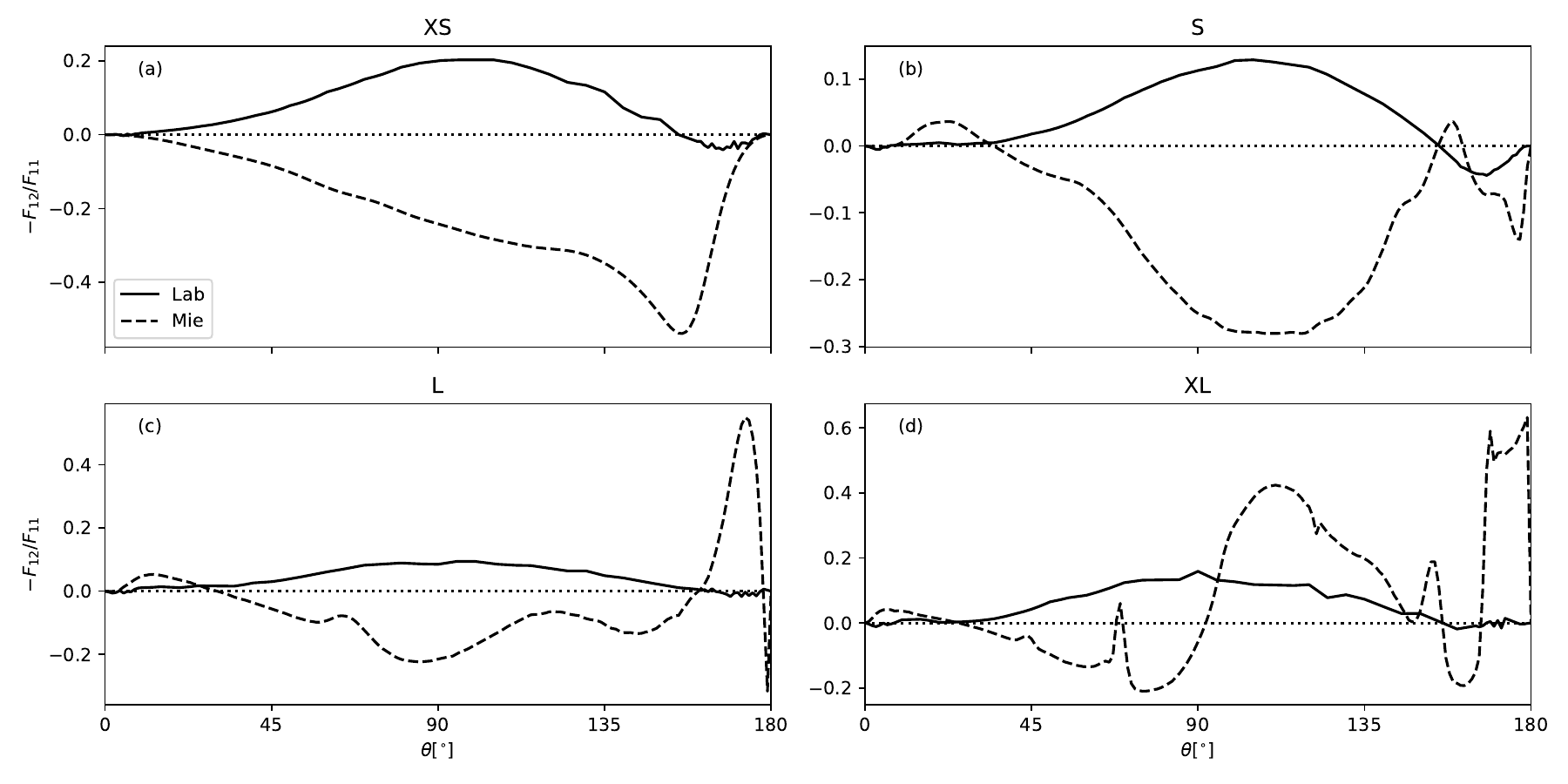}
    \caption{
        The degree of linear polarization (DLP) defined by $-F_{12} / F_{11}$ curves of the different laboratory samples (solid lines) compared to their corresponding Mie calculations (dashed lines). 
        }
    \label{fig:samples_mie}
\end{figure*}

Fig.~\ref{fig:samples_mie} compares the laboratory DLP with its respective Mie calculation (see Fig.~9 in \citealt{Munoz2021} for a similar plot). Across all four samples, the Mie calculations are drastically different from the laboratory measurements. Most notably, the Mie calculations do not follow a simple bell shaped curve with the peak at $\theta \sim 90^{\circ}$. Furthermore, the sign of the Mie DLP are mostly negative, which is the well-known polarization reversal \citep{Kataoka2015, Yang2016}.  In other words, the scattered light will be polarized in the scattering plane as opposed to being polarized perpendicular to the scattering plane for the Rayleigh regime or, evidently, the experimental DLP. The consequences of the differences between the Mie and experimental DLP's will be seen in the disk polarization images in Section \ref{sec:pol_ang} below. 

\begin{table}
    \centering
    \begin{tabular}{l|c|c|c|c}
        Sample & $x_{\text{eff}}$ & $a_{\text{eff}}$ [mm] & w \\
        \hline 
        XS & 4.4 & 0.70 & 0.9999 \\
        S & 17 & 2.7 & 0.9994 \\
        L & 43 & 6.8 & 0.9988 \\
        XL & 575 & 91.6 & 0.8710
    \end{tabular}
    \caption{
        The table lists the measured effective size parameter $x_{\text{eff}}$, the corresponding effective size $a_{\text{eff}}$ scaled to an observing wavelength of 1 mm, and the albedo for each sample. 
    }
    \label{tab:sample}
\end{table}

\subsection{Disk Model Setup} \label{ssec:disk_model_setup}

We consider a fiducial disk model with a dust surface density following a simple prescription \citep{Lyndenbell1974}:
\begin{align}
    \Sigma(R) = \Sigma_{c} \bigg( \dfrac{R}{ R_{c} } \bigg)^{-\gamma} 
        \exp{ \bigg[ - \bigg( \dfrac{R}{ R_{c}}  \bigg)^{2 - \gamma} \bigg] } , 
\end{align}
where $R$ is the cylindrical radius, $R_{c}$ is the characteristic radius, and $\gamma$ is the exponent that determines the radial power-law and exponential taper. The characteristic surface density is $\Sigma_{c} = \tau_{0}/\kext$ where we define $\tau_{0}$ as the characteristic optical depth in the vertical direction of the disk. The prescription allows us to scale the optical depth through $\tau_{0}$ directly since the $\kext$ is canceled out with opacity. We fix $R_{c}=50$ au as a representative size scale of dust disks \citep[e.g.][]{Andrews2020, Sheehan2022}. We set $\gamma=-0.2$ motivated by modeling of HL Tau \citep{Kwon2011, Kwon2015} and to connect to previous studies \citep{Yang2017}. The temperature is vertically isothermal and the radial profile goes as $R^{-0.5}$ with 30 K at 50 au to represent a passively heated disk \citep[e.g.][]{Chiang1997, Dullemond2001}. 

Motivated by vertical hydrostatic equilibrium and dust settling \citep[e.g.][]{Dubrulle1995}, the vertical dust density follows a Gaussian distribution 
\begin{align}
    \rho(R,z) = \dfrac{ \Sigma }{ \sqrt{2\pi} H } \exp{ \bigg[ - \dfrac{1}{2} \bigg( \dfrac{z}{ H } \bigg)^{2}  \bigg] }
\end{align}
where $H$ is the dust scale height as a function of radius. For simplicity, we adopt a power-law for the dust scale height 
\begin{align}
    H(R) = H_{0} \bigg( \dfrac{ R }{ R_{c} } \bigg)^{1.25}
\end{align}
where $H_{0}$ is the dust scale height at $R_{c}$. The prescription allows the freedom to study the effects of the geometrical thickness of the disk by changing $H_{0}$.

We use the three-dimensional Monte Carlo radiative transfer code, RADMC-3D to simulate the full Stokes ($I$, $Q$, $U$, $V$) images. The disk is set at a $45^{\circ}$ inclination and each image used $10^{9}$ photons.

\section{Results} \label{sec:results}

\subsection{Fiducial Model} \label{sec:pol_ang}
In this section, we compare the polarization images calculated from the laboratory samples and those from Mie theory. The linear polarized intensity $P$ is defined as $P \equiv \sqrt{Q^{2} + U^{2}}$, while the linear polarization fraction is $p_{f} \equiv P / I$. We first choose $H_{0}=0.5$ au and $\tau_{0}=1$ which represents a geometrically thin and optically thin disk. The chosen dust scale height is small since the millimeter emission of disks are generally observed to be thin, roughly $1\%$ of the radius \citep[e.g.][]{Pinte2016, Villenave2020}. The value of $\tau_{0}=1$ makes the interpretation simple for this section. We consider a larger optical depth in Section~\ref{sec:forward_scattering}. 

\begin{figure*}
    \centering
    \includegraphics[width=\textwidth]{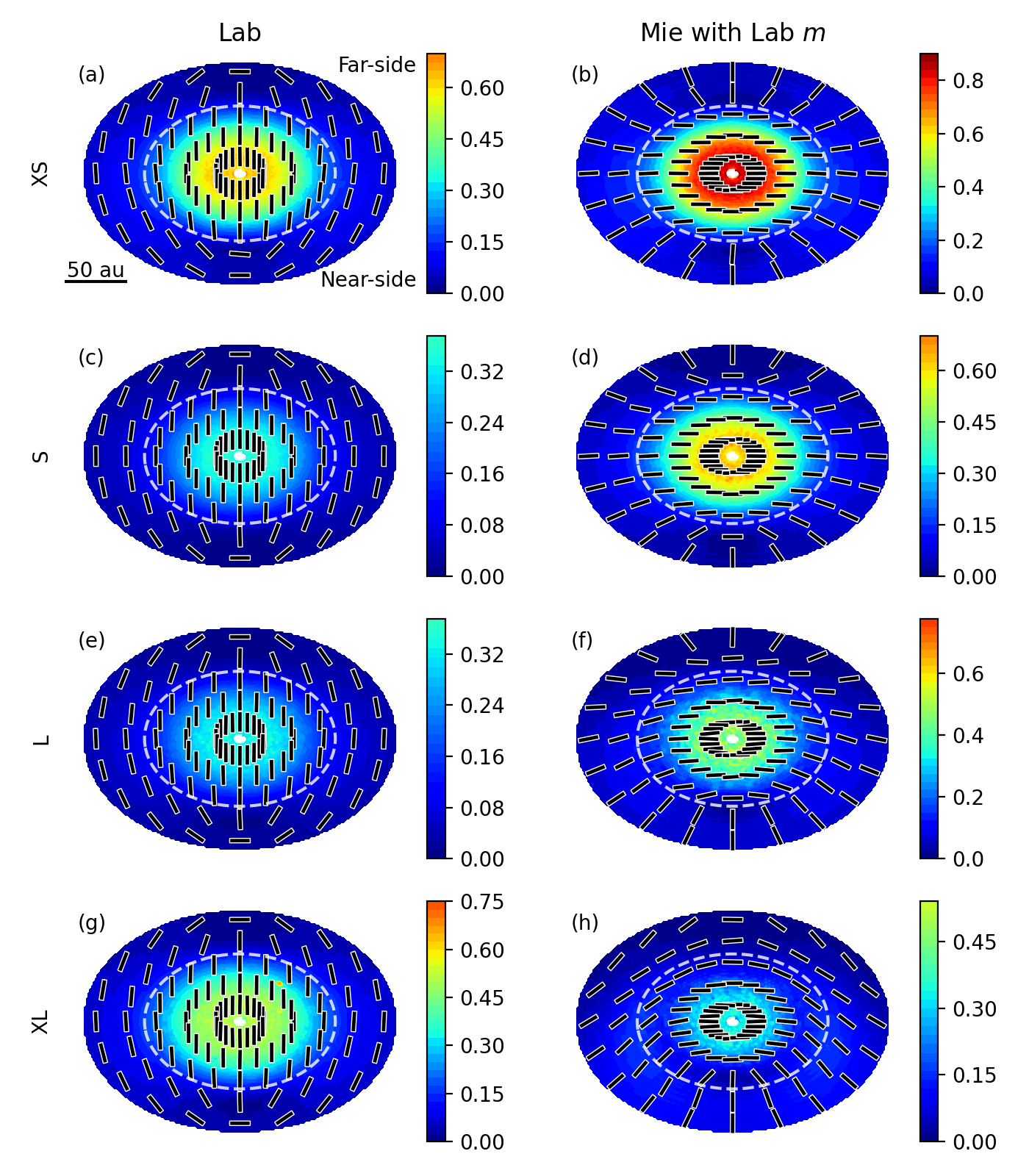}
    \caption{
        The images of linear polarization fraction, $p_{f}$, in percent (color maps) for different samples of experimental dust grains and their corresponding Mie calculations. The left column are produced from the experimental scattering matrix, while the right column are produced from Mie calculations. The top to the bottom row correspond to the XS, S, L, and XL samples. The polarization direction are denoted by the line segments. The dashed white contours are where the total optical depth is 0.1. The color scales are the same across images. 
    }
    \label{fig:disk_lpol}
\end{figure*}

The left column of Fig.~\ref{fig:disk_lpol} shows the polarization fraction and polarization direction images of the disk using the various samples of experimental dust grains. The relative levels of $p_{f}$ roughly correspond to the relative levels of their maximum DLP in Fig.~\ref{fig:samples_z11_dlp}b which is expected given the similar albedos ($\sim 1$ for the XS, S, and L sample and $\sim 0.87$ for the XL sample). The images show polarization that is parallel to the disk minor axis at the central regions (roughly within the region where the total optical depth is $\sim 0.1$ traced by the dashed white contour), while the outer regions of the disk show polarization that is more azimuthal. The two features are qualitatively similar to the pattern from Rayleigh scattering where the polarization parallel to the disk minor axis is simply due to inclination and the outer region is expected from radiation anisotropy \citep[e.g.][]{Kataoka2015, Yang2016}. This resemblance is expected because the DLP curves of Fig.~\ref{fig:samples_z11_dlp}b exhibit similar positive bell-shaped curves (scattered light is polarized perpendicular to the scattering plane) as the Rayleigh scattering approximation.

As a comparison, we show the polarization fraction images using the corresponding Mie theory calculations in the right column of Fig.~\ref{fig:disk_lpol}. The most striking result is the 90$^{\circ}$ offset in the polarization direction at the center of the disk between the lab scattering matrix and Mie calculations. This is true for all the samples considered here. The Mie calculations produce negative DLP (Fig.~\ref{fig:samples_z11_dlp}) for these large grains which causes the polarization to become parallel to the disk major axis. Between the Lab and Mie model images for samples XS, S, and L, the outer regions of the disk are perpendicular to each other also because of the opposite sign in the DLP. The outer regions of the Lab and Mie XL sample models are not entirely perpendicular to each other because the DLP of the Mie calculation (Fig.~\ref{fig:samples_z11_dlp}d) is mostly opposite when $\theta$ is less than $\sim 90^{\circ}$ (the forward scattering half) while it has the same sign when $\theta$ is greater than $\sim 90^{\circ}$ (the back scattering half). Since radiation mostly travels outward for the outer region, the polarization direction is mostly perpendicular to the Lab counterpart at the near side (bottom half of the image) where most photons are forward scattered, while polarization is mostly parallel to the Lab counterpart at the far side (upper half of the image) where most photons are back scattered.

The level of polarization fraction from Mie calculations also do not resemble the corresponding laboratory samples. The level of polarization of images from laboratory matrices are generally lower than its Mie counterpart. The XS and XL samples have $\sim 0.5\%$ peak polarization, while the S and L samples have $\sim 0.3 \%$. The images from Mie calculation have peak levels of polarization at $\sim 0.8\%$ with the XS sample and at $\sim 0.4\%$ with the XL sample. The largest grain sizes have the lowest level of polarization, which is different from the images using laboratory matrices.

\subsection{Effects of Strong Forward Scattering} \label{sec:forward_scattering}

The most striking feature from Section~\ref{sec:pol_ang} is the difference in the polarization angle between the calculations using lab measurements and Mie results since the DLP for the large irregular grains in consideration is more similar to the Rayleigh scattering behavior. However, if grains are indeed large, we would expect large forward scattering which is drastically different from the Rayleigh limit (Fig.~\ref{fig:samples_z11_dlp}). 

In the limit where photons travel radially in the midplane of the disk, we would expect photons at the near side of the inclined disk to be forward scattered to reach the observer as opposed to the far side where photons will be more back scattered. Given that the forward scattering peak of the phase function $F_{11}$ is a few orders of magnitude larger than side-scattering or back-scattering, one may expect scattering by large irregular grains to potentially cause a significant near-far side asymmetry. In this section, we show that the near-far side asymmetry can be significant if the scattering dust disk is geometrically thick, but the asymmetry almost disappears for a geometrically thin disk. As we explain below, this is an extension from the near-far side asymmetry when the disk is geometrically thick \textit{and} optically thick as demonstrated in \cite{Yang2017} when grains do not have large forward scattering peaks.

\begin{figure*}
    \centering
    \includegraphics[width=\textwidth]{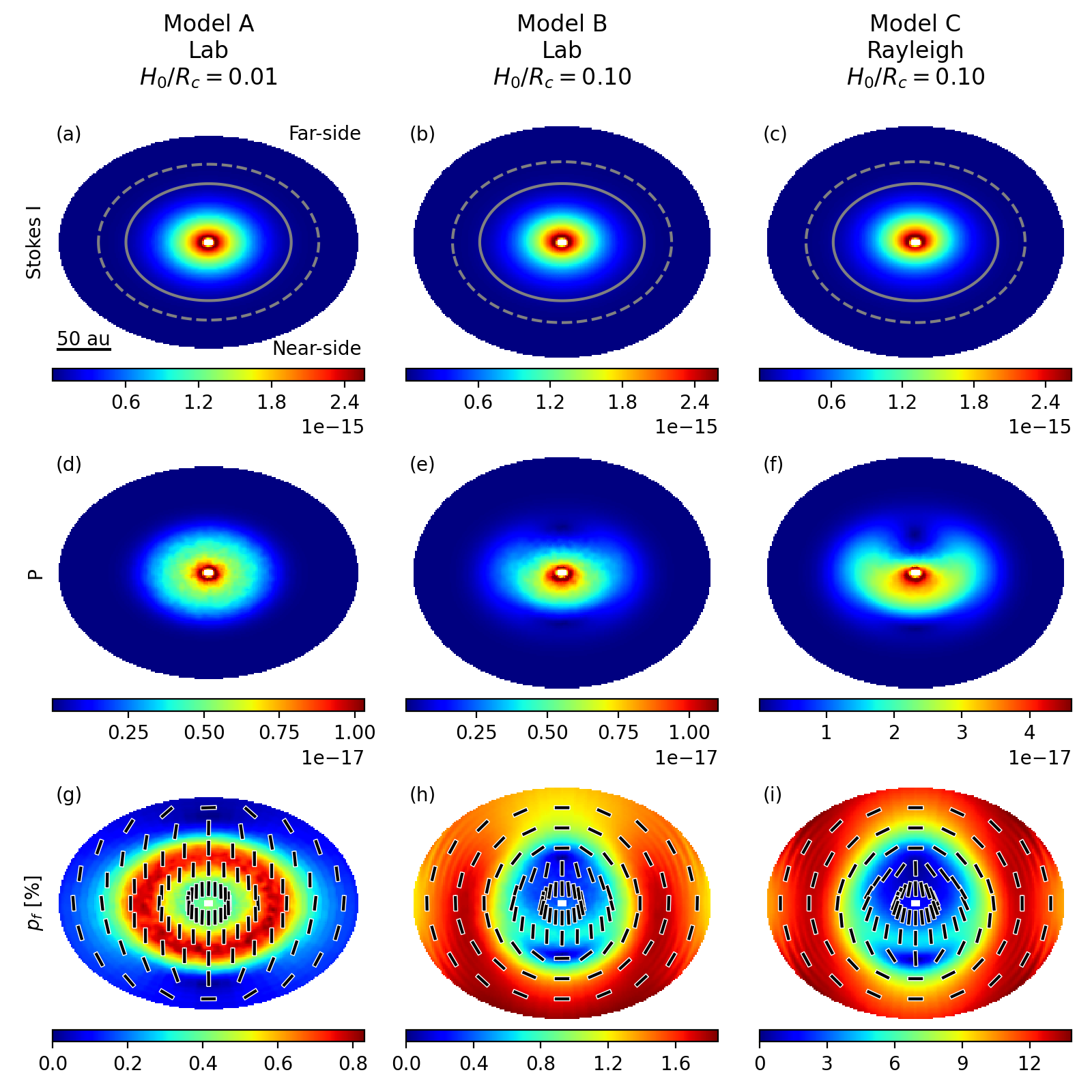}
    \caption{
        The polarization images for Models A, B, and C (left to right). The color map of the top row is the Stokes $I$, while the dashed and solid grey contours mark where the optical depth is 0.1 and 1. The second row is $P$. Stokes $I$ and $P$ are both in erg s$^{-1}$ sr$^{-1}$ cm$^{-2}$ Hz$^{-1}$. The third row is $p_{f}$ in percent with the polarization direction denoted as vectors. 
    }
    \label{fig:H_opac_image}
\end{figure*}

Since the phase function and DLP are similar for all the dust samples, we only use the XS sample for illustration. We set $\tau_{0}=10$ to increase the optical depth of the disk which allows us to compare the optically thick region near the center versus the optically thin part at larger radius. Fig.~\ref{fig:H_opac_image} shows the Stokes $I$, polarized intensity $P$, and polarization fraction $p_{f}$ for three different cases described below. We also plot the optical depth along the line of sight as contours in the top row of Fig.~\ref{fig:H_opac_image} to help diagnose the images. The names of each model and the corresponding parameters are listed in Table~\ref{tab:forward_scattering_models}. 

\begin{table}
    \centering
    \begin{tabular}{c|c|c}
        Name & $H_{0}$ & Scattering matrix \\
        \hline
        Model A & 0.5 au & XS \\
        Model B & 5 au & XS \\
        Model C & 5 au & Rayleigh
    \end{tabular}
    \caption{
        Column (1): The names of the models used in Section~\ref{sec:forward_scattering}. Column (2): the value for the dust scale height $H_{0}$. Column (3): The scattering matrix sample. 
    }
    \label{tab:forward_scattering_models}
\end{table}

The left column of Fig.~\ref{fig:H_opac_image} is a model with $H_{0}=0.5$ au which corresponds to a geometrically thin disk with $H_{0}/R_{c}=0.01$ (Model A). The $p_{f}$ of the geometrically thin disk in Fig.~\ref{fig:H_opac_image}g is the optically thick counterpart of Fig.~\ref{fig:disk_lpol}a. A notable difference between Fig.~\ref{fig:H_opac_image}g and Fig.~\ref{fig:disk_lpol}a is the peak of $p_{f}$ forms a ring for the optically thick case (Fig.~\ref{fig:H_opac_image}g), but $p_{f}$ peaks in the center of the image for the optically thin case (Fig.~\ref{fig:disk_lpol}a). This is because $p_{f}$ peaks where the optical depth along the line of sight is of order unity \citep{Yang2017} which is seen in Fig.~\ref{fig:H_opac_image}a. Evidently, the Stokes $I$, polarized intensity, and $p_{f}$ do not have obvious near-far side asymmetry. The lack of obvious asymmetry is perhaps not surprising because the radiation is roughly isotropic in the midplane for this geometrically thin disk. In other words, there are a comparable number of photons that scatter at all angles to the observer for grains either in the near side or the far side.

Increasing the dust scale height can increase the radiation anisotropy \citep{Ohashi2019}. Thus, in the middle column of Fig.~\ref{fig:H_opac_image}, we consider $H_{0}=5$ au which is an increase of dust scale height by a factor of 10 (Model B). One can easily identify differences between the near side and far side for at least $P$ and $p_{f}$. The near side of $P$ (Fig.~\ref{fig:H_opac_image}e) just below the center is brighter than the far side. The brightest part of $P$ resembles a ``kidney'' which also appears in \cite{Yang2017}. For $p_{f}$ in Fig.~\ref{fig:H_opac_image}h, there is a horizontal bar of $\sim 0.5\%$ (with vertical polarization) at the near side just outside the center. Additionally, the $p_{f}$ at the edge of the disk for the near side (with horizontal polarization) is clearly larger than $p_{f}$ (which is also horizontally polarized) at the far side.

As a comparison, we consider the same geometrically thick disk, but we use the scattering matrix in the Rayleigh limit (Model C) while adopting the same albedo and $\kext$ as the previous cases. Similar to the Model B, the Rayleigh limit counterpart shown in the right column of Fig.~\ref{fig:H_opac_image} also shows clear near-far side asymmetry. The central regions of $P$ in Fig.~\ref{fig:H_opac_image}f also looks like a ``kidney'' overall with the near side being brighter than the far side. In Fig.~\ref{fig:H_opac_image}i, the $p_{f}$ also has a horizontal bar of vertical polarization immediately outside of the center region at the near side. However, the near and far sides of the outer region with horizontal polarization appear symmetric, which is in contrast to  those in Fig.~\ref{fig:H_opac_image}h. The larger levels of $P$ and $p_{f}$ (up to $\sim 12\%$) across the image is because the peak DLP in the Rayleigh limit is larger than the peak DLP of sample XS (Fig.~\ref{fig:samples_z11_dlp}b). The peak DLP in the Rayleigh limit is $\sim 5$ times the peak DLP of sample XS which explains the difference between the $p_{f}$ images.

\begin{figure}
    \centering
    \includegraphics[width=\columnwidth]{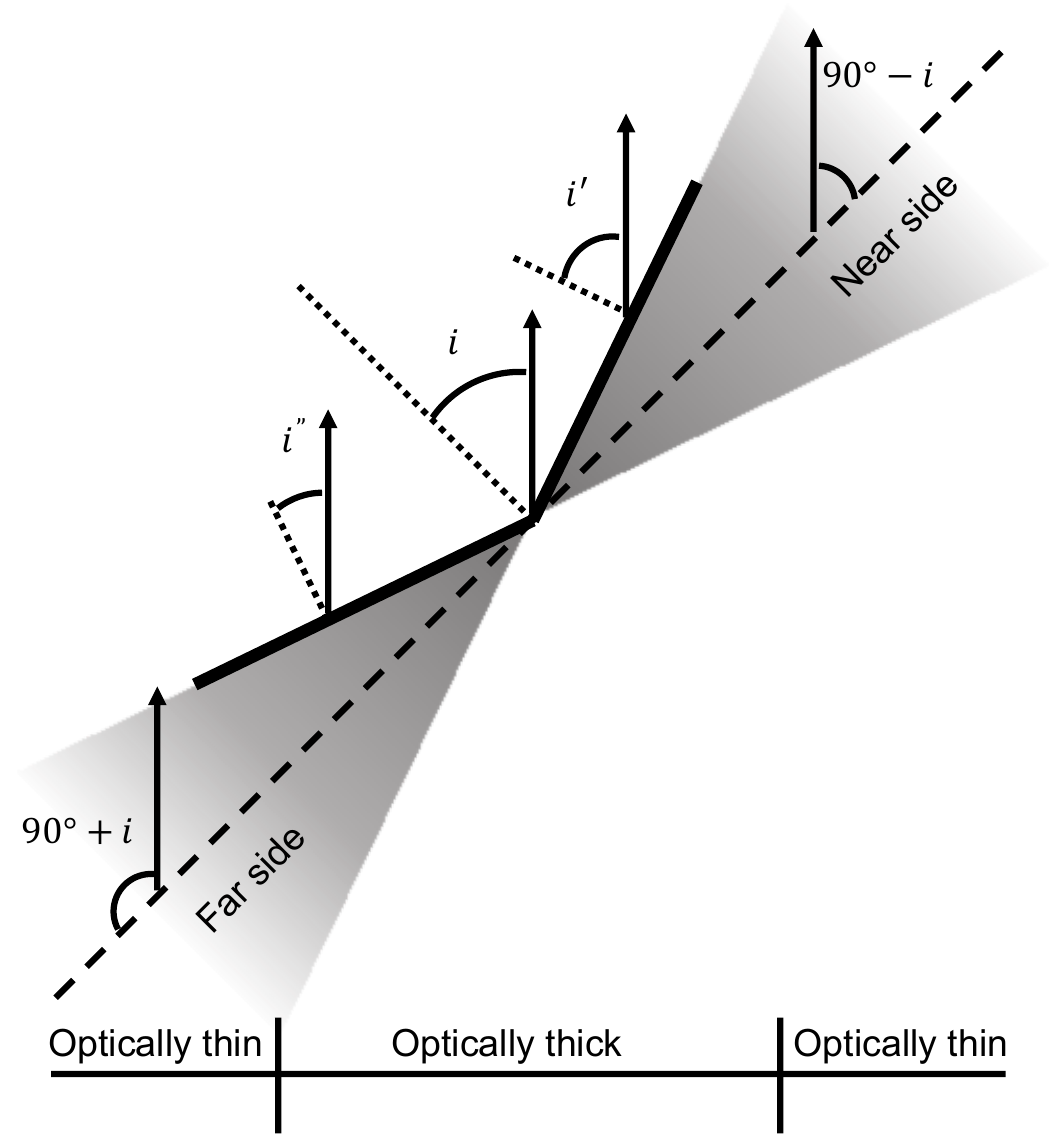}
    \caption{
        A schematic diagram of a cross section of the disk and its relation to the line of sight. The observer is viewing the disk from the top of the diagram. The arrows represent the direction to the observer. The disk midplane is the dashed line and the plane is inclined by $i$ with respect to the plane of sky. The near side and far side are labeled. In the optically thin region, the scattering angle for the near side is $90^{\circ}-i$, while that for the far side is $90^{\circ}+i$. In the optically thick region, the surface of the disk forms an effective local inclination to the observer. The local inclination of the near side $i'$ is larger than the that of the far side $i''$.  
    }
    \label{fig:schem}
\end{figure}

The polarization of the outer region of Model B is in fact due to what we expect from strong forward scattering from radiation anisotropy: the radiation from the central regions of the disk propagates to the edge and is scattered to the observer \citep{Kataoka2015}. With strong forward scattering, the near side of the disk scatters more of the polarized photons to the observer. Fig.~\ref{fig:schem} is a schematic diagram of a meridional cross section of the disk (which is an extension of Fig.~6 from \citealt{Yang2017}). In the optically thin regime (large radiation anisotropy), the angle between the radial direction along the midplane to the observer is $90^{\circ}-i$ for the near side. On the other hand, the far side scatters by $90^{\circ}+i$. Given that $F_{11}$ in the Rayleigh limit does not have a forward scattering peak and is, in fact, symmetric from $90^{\circ}$, the near-far-side asymmetry disappears in the outer optically thin region. 

Both Model B and C exhibit larger $P$ and $p_{f}$ at the near side at least near the center. This is due to the disk surface effects demonstrated in \cite{Yang2017}. The polarization increases with increasing inclination of the surface if the line of sight is optically thick \citep{Yang2017}. Since the local surface of the near side is more inclined than the local surface of the far side as illustrated in Fig.~\ref{fig:schem}, the $p_{f}$ of the near side is higher.

\begin{figure*}
    \centering
    \includegraphics[width=\textwidth]{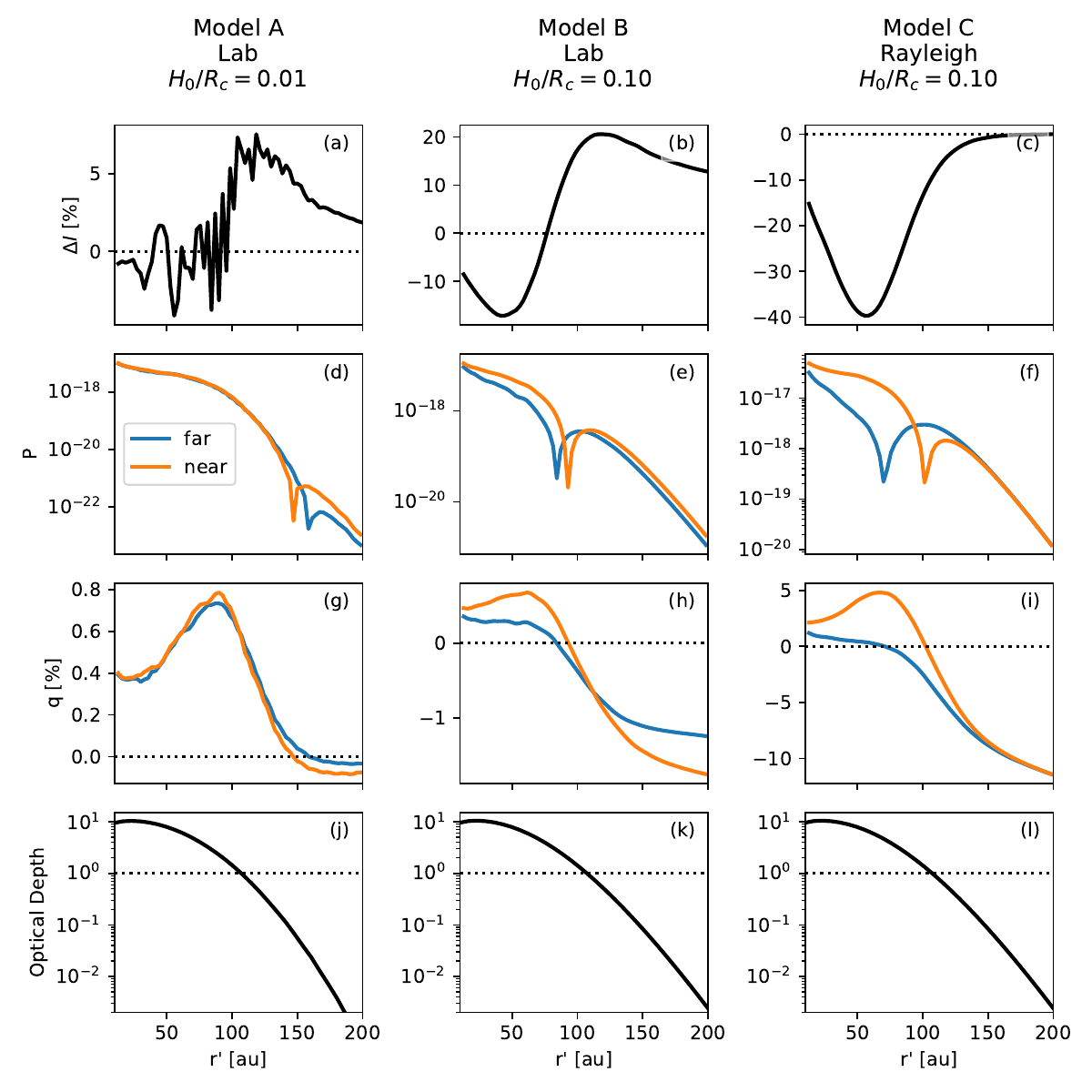}
    \caption{
        Cuts along the minor axis comparing the near side and the far side as a function of the deprojected radius $r'$ in au. The columns from the left to right are Models A, B, and C respectively. The rows from the top to bottom are $\Delta I$ in percent, $P$ in erg s$^{-1}$ sr$^{-1}$ cm$^{-2}$ Hz$^{-1}$, $q$ in percent, and the total optical depth. Since the optical depth is symmetric across the disk major axis, only one line is plotted. For $P$ and $q$, the near side is plotted in orange and the far side is plotted in blue. 
    }
    \label{fig:H_opac_nearfar}
\end{figure*}

To examine the near-far side asymmetry in more detail, we make cuts along the minor axis and compare the near-far side profiles as a function of distance from the center. Since Stokes $I$ is fairly similar across the near and far side, we plot the relative difference of Stokes $I$ between the near side and far side defined as 
\begin{align}
    \Delta I \equiv \dfrac{ I_{\text{near}} - I_{\text{far}} }{ 0.5 ( I_{\text{near}} + I_{\text{far}}) }
\end{align}
where $I_{\text{near}}$ and $I_{\text{far}}$ are the Stokes $I$ for the near side and far side respectively in the top row of Fig.~\ref{fig:H_opac_nearfar}. The second row of Fig.~\ref{fig:H_opac_nearfar} shows the linear polarized intensity. The third row shows the polarization fraction, but we use $q\equiv Q/I$ because Stokes $U=0$ along the disk minor axis due to the symmetry and $q$ completely describes the polarization fraction. Using $q$ is convenient because the sign gives the polarization direction: positive $q$ means vertical polarization or polarization parallel to the disk minor axis for our setup and negative $q$ means polarization perpendicular to the disk minor axis. The last row of Fig.~\ref{fig:H_opac_nearfar} shows the total optical depth along the line of sight $\tau$. Note that there is only one curve because $\tau$ is symmetric across the major axis for an axisymmetric disk. For clarity, we use $\tau'$ to denote the optical depth from a particular location along the line of sight to the observer. 

For the geometrically thin Model A, the images in the left column of Fig.~\ref{fig:H_opac_image} show little near-far side asymmetry, and indeed the left column of Fig.~\ref{fig:H_opac_nearfar} shows little difference between the near-far side. The more noticeable asymmetry is its $P$ and $q$ at the edge of the disk when $\tau< 1$. The asymmetry is similar to the geometrically thick case (Model B) which can be easily identified and understood first. 

In Fig.~\ref{fig:H_opac_nearfar}b, the positive $\Delta I$ at larger radius means the near side is brighter than the far side which is less visible in the image (Fig.~\ref{fig:H_opac_image}b). The positive $\Delta I$ is expected because, in the regime where radiation anisotropy is in the radial direction, photons from the near side are more forward scattered than the those from the far side. In contrast, $\Delta I$ of Model C (Fig.~\ref{fig:H_opac_nearfar}c) is near 0 at larger radius because there is no strong forward scattering. 

The negative $\Delta I$ at smaller radius in Fig.~\ref{fig:H_opac_nearfar}b means the far side is brighter than the near side for Model B. As discussed in \cite{Yang2017}, this is because for the same projected distance from the center, the line of sight of the far side has its $\tau'=1$ surface is at a smaller radius with higher temperature than the line of sight of the near side (also depicted in Fig.~\ref{fig:schem}). The behavior is similar to Model C (Fig.~\ref{fig:H_opac_nearfar}c) even though $F_{11}$ is completely different. This is expected since the radiation field is more isotropic and thereby weakens the effects of differences in $F_{11}$. Interestingly, for Model B, the far side is brighter than the near side only by $\sim 10\%$, whereas for Model C, the far side is brighter by $\sim 20\%$. It appears that although the local surface effect dominates, forward scattering still provides some extra boost in the near side and counteracts the near-far side asymmetry from local surface effects alone.

The linear polarized intensity of Model B in Fig.~\ref{fig:H_opac_nearfar}e also has a slightly brighter near side at large radius when $\tau < 1$ ($\sim 1.5$ times brighter that the far side). At the same time, in Fig.~\ref{fig:H_opac_nearfar}h, $q$ of the near side is more negative than $q$ of the far side. Both are due to forward scattering: the strong forward scattering peak consistently provides an extra amount of scattered photons that are horizontally polarized and results in an increased linear polarized intensity and extra negative $q$ for the near side. In contrast, for the Rayleigh limit case (Model C), the linear polarized intensity in Fig.~\ref{fig:H_opac_nearfar}f and $q$ in Fig.~\ref{fig:H_opac_nearfar}i are equal across the near side and far side as expected for large radius. At smaller radius when $\tau> 1$, both Models B and C show similar behaviors in $P$ and $q$ also due to a more isotropic radiation field. The larger $P$ and $q$ in the near side for both models are simply due to the location surface effect mentioned above and illustrated in Fig.~\ref{fig:schem}.

Returning to Model A, we can identify similar features like those in Model B, but with reduced levels of asymmetry. At larger radius ($\tau \lesssim 1$), $\Delta I$ is positive which indicates a brighter near side due to forward scattering (Fig.~\ref{fig:H_opac_nearfar}a), but it is only at $\sim 5\%$ because the radiation anisotropy is much smaller than Model B. Likewise, the polarization of the near side is stronger (Fig.~\ref{fig:H_opac_nearfar}d) and horizontal (Fig.~\ref{fig:H_opac_nearfar}g). At smaller radius ($\tau > \sim 1$), $\Delta I$ is essentially zero and polarization is equal across the near and far side because the geometrically thin disk suppresses local surface effects like those shown in Model B. In other words, the disk is essentially what we would expect from an infinitely flat disk.

\subsection{Effects of the refractive index} \label{sec:refractive_index}

The refractive index of $1.65 + 10^{-5} i$ in Section~\ref{sec:results} was for forsterite at the experiment wavelength of $514$ nm. Although the composition of grains in protoplanetary disks is unclear, the commonly adopted material for disks, such as astronomical silicates, are usually more absorptive at (sub)millimeter wavelengths \citep[e.g.][]{Draine2003, Birnstiel2018}. For example, the commonly used opacity mixture from The Disk Substructures at High Angular Resolution Project (DSHARP) has a refractive index of $\sim 2.30 + 0.021 i$ at $\lambda=$1mm \citep{Birnstiel2018}. Hypothetically, if the material in disks can indeed have an imaginary part of the refractive index $k\sim 10^{-5}$ at millimeter wavelengths, then we can expect the irregular grains to produce $\sim 0.5\%$ polarization even for grains in the geometric optics regime (XL sample) as demonstrated in Fig.~\ref{fig:disk_lpol}. However, in the conventional scenario with more absorptive material, there are two key effects that can change the interpretation of large grains: the peak of the DLP and the albedo.

First, the shape of the DLP should remain bell-shaped with an increase in the peak level. Simulations done by \cite{Shen2009} used absorptive (e.g., silicates; \citealt{Draine2003_silicates}) and compact aggregates with size parameters greater than unity and are able to reproduce bell-shaped DLP's similar to the experimentally derived scattering matrix from forsterite (see also \citealt{Zubko2009} and \citealt{Tazaki2019}). Laboratory results also show that large irregular grains of more absorptive material maintains the bell-shaped DLP, with a {\it higher} peak of the DLP (see e.g., \citealt{Munoz2007}, \citealt{Frattin2019}). Thus, we can expect that irregular dust with refractive index closer to that of silicates at (sub)millimeter wavelengths would give the same qualitative results as shown in Section~\ref{fig:disk_lpol} and perhaps further increase the level of polarization for each of the sample. That increase would make it easier to match the observations of $\sim 1\%$ polarization.

However, the level of polarization is complicated by the other contributing factor which is the albedo. The albedos from Mie theory used for the forsterite samples were $\sim 1$ which is expected given the very low $k$. In the limit of $k=0$, meaning no absorption, $\kabs=0$ and $w=1$; \citealt{van_de_Hulst1957}) and this is true for all the samples considered here. When using the commonly adopted DSHARP refractive index of $m=2.30 + 0.021 i$, the albedos for the XS, S, L, and XL samples are $\sim 0.80$, $0.66$, $0.59$, and $0.12$ respectively. The decrease in albedo with increasing grain size is due to stronger forward scattering which are truncated within $1^{\circ}$ (see Section~\ref{sec:dustmodel}). One can see this from the albedos before truncation which are $\sim 0.80$, $0.67$, $0.60$, and $0.60$ and evidently do not change much with increasing grain size (see also \citealt{Kataoka2015, Tazaki2019}). Thus, for the absorptive material, $\kabs$ can become comparable to the effective scattering opacity or even dominate like for the XL sample and we expect the resulting level of polarization to decrease.

To illustrate the effects of the differences in albedo due to increased $k$. we follow the same procedure as described in Section~\ref{sec:pol_ang} to produce the disk images, but adopt the new set of albedo. Note that the experimental scattering matrix is kept unchanged, since we expect the same qualitative behavior and only moderate quantitative differences caused by the increase of $k$.

\begin{figure}
    \centering
    \includegraphics[width=\columnwidth]{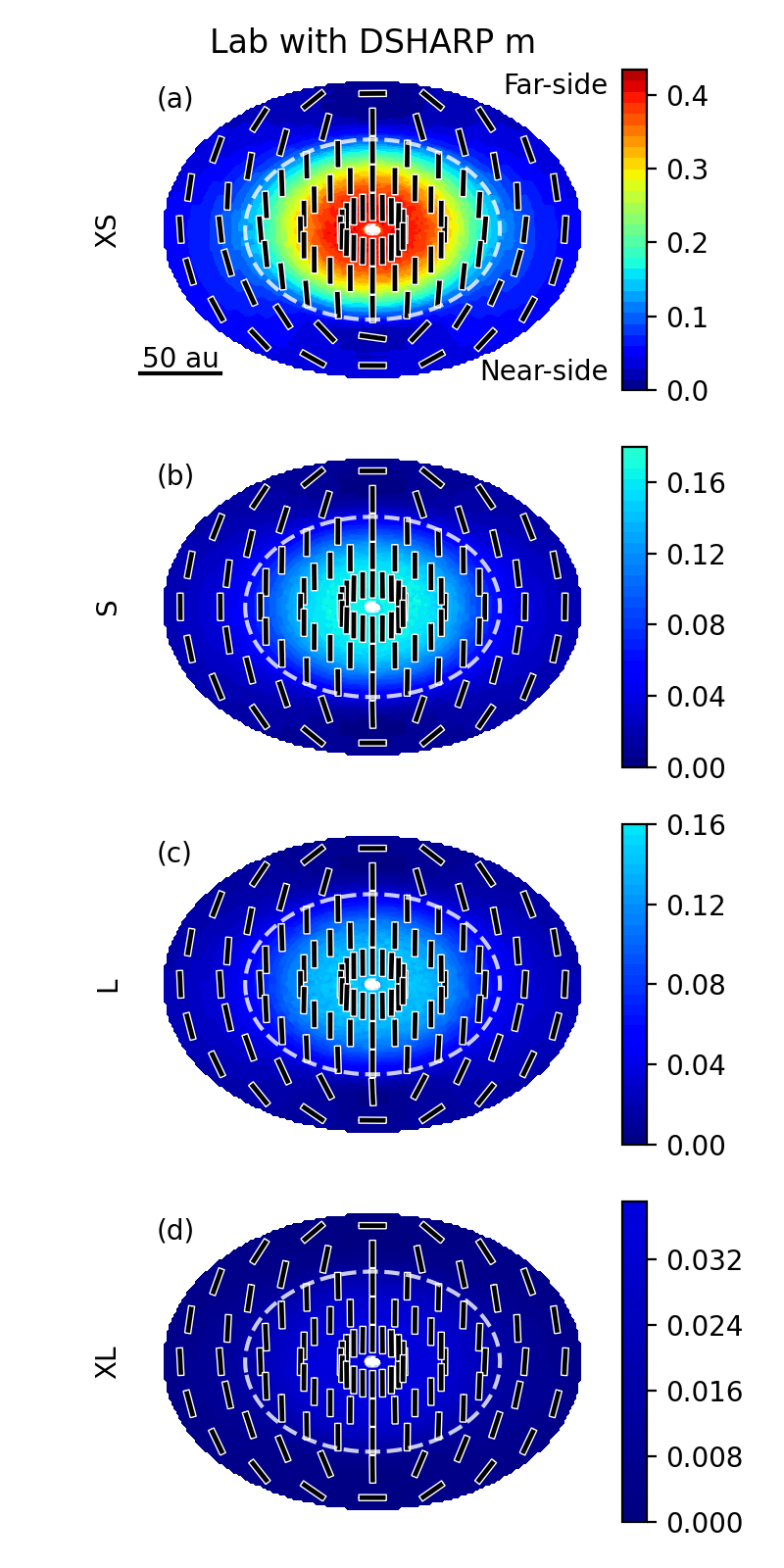}
    \caption{
        The images of the polarization percentage $p_{f}$ using the DSHARP mixture which is $m=2.30 + 0.021 i$ at $\lambda=1$mm. The figure is plotted in the same way as and should be compared to the left column of Fig.~\ref{fig:disk_lpol}. The polarization directions remain parallel to the disk minor axis, but the level of polarization is decreased across all samples due to the lower albedo especially for the larger samples.  
    }
    \label{fig:disk_lpol_dsonk}
\end{figure}

Fig.~\ref{fig:disk_lpol_dsonk} shows the polarization images which should be compared to the left column of Fig.~\ref{fig:disk_lpol}. As expected, all samples produce polarization directions that are parallel to the disk minor axis (since this only depends on the DLP which was kept the same), but given the decrease in albedo, $p_{l}$ for each sample is consistently lower compared to its forsterite counterpart. While $p_{l}$ for the XS sample can reach $\sim 0.4\%$, $p_{l}$ for the XL sample reaches only $\sim 0.03\%$. The actual polarization levels are expected to be somewhat higher (potentially reaching percent-level for the XS sample) because, as mentioned earlier, more absorptive materials have higher peak values of the DLP than that of the forsterite sample adopted in Fig.~\ref{fig:disk_lpol_dsonk} (see e.g., \citealt{Munoz2007}, \citealt{Frattin2019}). Nevertheless, very large size parameters, like the XL sample, are unlikely to produce observable polarization percentages even with the increased DLP. This essentially demonstrates that using the larger $k$ decreases the acceptable range of $x$ as the albedo becomes too low for very large $x$.

In contrast, the DLP of spherical grains typically used for inferring maximum grain sizes, quickly becomes negative after $x$ becomes greater than of order unity, while the albedo decreases more slowly \citep[e.g][]{Kataoka2015, Tazaki2019, Yang2020}. As an example, the Mie calculations for the smallest XS sample ($x_{\text{eff}}=4.4$) using $m=2.30 + 0.021i$ gives a DLP at $\theta=90^{\circ}$ of $\sim -0.16$ which is already negative, while the albedo of $\sim 0.8$ is still relatively high. Thus, consistent with the conclusion from Section~\ref{sec:pol_ang}, the irregularity of large grains remains a possible mechanism to remove the strict upper limit of $x\sim1$ from spherical grains. How large the grains can be depends, in part, on $k$. Accounting for more absorptive material, the XS sample remains likely to produce a correct polarization direction and a detectable level of polarization at millimeter wavelengths through scattering. 


\section{Discussion} \label{sec:discussion}

\subsection{Tensions of grain size compared to the opacity index} \label{sec:tensions_grain_size}
Many sources have continuum linear polarization level that is $\sim 1\%$ at Band~7 ($870 \mu$m) of ALMA and polarization is parallel to the disk minor axis \citep[e.g.][]{Stephens2017, Bacciotti2018, Cox2018, Hull2018, Dent2019, Mori2019}. The pattern is best explained by scattering of spherical grains with size parameters of order unity (usually assuming a power-law size distribution up to some maximum grain size and refractive indices similar to the DSHARP mixture; e.g., \citealt{Kataoka2016, Tazaki2019, Yang2020}). Thus, it appears that several disks are fine tuned to have maximum grain sizes of $\sim 100 \mu$m.

Our results demonstrate that irregular grains alleviate the need for the $\sim 100 \mu$m grains to explain the polarization angle. The strict upper limit of grain size inferred from polarization is due to the assumption of perfectly spherical grains which causes the DLP to quickly become negative when $x$ is of order unity (see e.g. \citealt{Yang2020} for a demonstration). As shown in Fig.~\ref{fig:samples_mie}, irregular grains with size parameters beyond 1 and even up to 525 as for sample XL can still maintain a well-behaving polarization curve which is bell-shaped and with little to no polarization reversal. The resulting images in Fig.~\ref{fig:disk_lpol} show that these large grains can produce polarization that is parallel to the disk minor axis as compared to the Mie calculations. However, accounting for realistic (sub)millimeter refractive index as explored in Section~\ref{sec:refractive_index}, the level of polarization observed by ALMA can rule out very large grains given its low albedo that diminishes the polarization level. The sample XS appears to be within the acceptable range and corresponds to large 0.7 mm grains at $\lambda=1$~mm whose existence has also been implied from the SED after accounting for optical depth effects and scattering \citep[e.g.][]{Carrasco2019, Sierra2021}.

A similar conclusion was demonstrated numerically by \cite{Tazaki2019} who calculated the scattering matrix of dust aggregates instead of solid spheres (\citealt{Tazaki2016, Tazaki2018}; see also \citealt{Kirchschlager2014}). The resulting DLP at $\theta=90^\circ$ at $\lambda=1$mm does not become negative even if the maximum grain size is $100$ cm (or a maximum size parameter of $\sim 6000$) approximated by the effective medium theory \citep{Kataoka2014}. A simulated disk image from \cite{Tazaki2019} also showed near-far side asymmetry in the polarized intensity and polarization fraction similar to Fig.~\ref{fig:H_opac_nearfar}. The similarity of the resulting images from both the experimental and the simulated scattering matrices for irregular grains strengthens the possibility that the observed disk polarization can be explained by scattering of large mm-sized grains.

From Fig.~\ref{fig:disk_lpol}, we have demonstrated that adopting perfectly spherical grains can lead to drastically incorrect predictions to (sub)millimeter polarization images of disks. Given the increasing number of polarization images from ALMA and fundamental importance of grain size, there is a strong need to improve upon the polarization predictions to make the most out of the hard-fought data. An obvious method, as demonstrated in this paper, is to increase laboratory measurements. In particular, the field will benefit from measuring scattering matrices at (sub)millimeter wavelengths with materials that match the grains in protoplanetary disks as closely as possible.

As mentioned earlier, \cite{Tazaki2019} showed that large porous grains can in principle produce polarization with a pattern and degree consistent with observations. This is particularly true for mm-sized grains with a porosity of $f=0.1$ (see the middle panel of their Fig.~6). In this paper, we have demonstrated that irregular mm-sized grains can do the same without any porosity (i.e., $f=1$). An advantage of nonporous mm-sized grains over their porous counterparts is that they are less well coupled aerodynamically to the gas in the disk, making them easier to settle to the midplane and participate in planet formation through, e.g., streaming instability and/or pebble accretion.



\section{Conclusions} \label{sec:conclusion}
ALMA has consistently detected (sub)millimeter polarization that is $\sim 1\%$ and parallel to the disk minor axis for many sources. This common polarization pattern has been interpreted as evidence for scattering by $\sim 100\mu$m-sized grains \citep[e.g.][]{Yang2016, Kataoka2016}, yet the opacity index $\beta$ suggests mm/cm grain sizes \citep[e.g.][]{Draine2006}. In this paper, we demonstrate that the $\sim 100 \mu$m sized grains inferred based on polarization is due to the strict assumption of spherical grains. We use realistic scattering matrices measured from the laboratory for irregular grains with size parameters ranging from 4 to 575 (corresponding to mm/cm-sized and even decimeter-sized grains for an observing wavelength of 1~mm) to simulate disk polarization images. Our results are as follows:
\begin{enumerate}[label=\arabic*)]
    \item The degree of linear polarization (DLP) for large irregular grains (much larger than the wavelength) derived from laboratory measurements remain mostly positive, i.e., the polarization of scattered light is perpendicular to the scattering plane for incoming non-polarized light. This is similar to Rayleigh scattering except with a maximum DLP that is $10 \sim 20\%$ for all the size parameters considered. In contrast, Mie calculations using matching large spherical grains produce DLP that is negative. As a result, the experimental scattering matrices for all the samples produce disk polarization that is parallel to the disk minor axis, whereas the Mie scattering matrices produce disk polarization that is parallel to the disk major axis. A major reason for the inferred $\sim 100 \mu$m size for spherical grains comes from the DLP becoming negative once the size parameter exceeds unity. After accounting for the more commonly adopted refractory index (such as that adopted for the DSHARP project), the level of polarization for irregular grains much larger than mm would be much lower than what is observed, while the grains with a size parameter of several (corresponding to mm-sized grains) remains likely to produce the observed polarization. 
    
    \item Since large grains produce a strong forward scattering peak, we find that forward scattering can create near-far side asymmetry in the disk image if the dust layer is not too geometrically thin and the disk is inclined to the line of sight. In the optically thin regions where most of the radiation travels radially outwards, the photons are more forward scattered at the near side of the disk and more backward scattered at the far side. As a result, the near side has boosted scattering with polarization that is parallel to the disk major axis. The polarization cancels with the polarization induced by inclination which is parallel to the disk minor axis. In the optically thick regions where the radiation is mostly isotropic, the role of forward scattering is minimal and the polarization fraction is larger at the near side because the local disk surface is more inclined at the near side. The degree of this near-far side asymmetry depends on forward scattering, which is a hallmark of scattering by large grains and can be used to infer their presence. 
\end{enumerate}

\section*{Acknowledgements}

ZYDL acknowledges support from the Jefferson Scholars Foundation, NASA 80NSSC18K1095, and also support from the ALMA Student Observing Support (SOS). ZYL is supported in part by NASA 80NSSC20K0533 and NSF AST-1910106. LWL acknowledges support from NSF AST-1910364. Research by OM is supported by grant PID2021-123370OB-I00 funded MCIN/AEI/FEDER, UE. We thank the anonymous reviewer for the constructive comments that improved the paper.

\section*{Data Availability}

The experimental scattering-matrix elements as functions of the scattering angle and the size distributions are freely available at the Granada–Amsterdam light-scattering database (\url{www.iaa.es/scattering}). Also see \cite{Munoz2021} for details of the data. Additional data underlying this article are available from the corresponding author upon request.



\bibliographystyle{mnras}
\bibliography{main} 




\appendix

\section{Comparing Truncation Angles} \label{sec:compare_truncation}

As described in Section~\ref{sec:dustmodel}, we adopted a truncation angle in the scattering matrix when producing the results from Monte Carlo radiative transfer. To understand the effects of truncating the forward scattering peak, we use the disk model from Section~\ref{sec:pol_ang} and the phase function of the XL sample which has the strongest forward scattering. The goal is to ensure that truncating the peak allows smoother polarization images with achievable number of photons without altering the quantitative results much. We consider three models with truncation angles $\theta_{c}$ at $1^{\circ}$, $2^{\circ}$, and $4^{\circ}$ and one model without any truncation, i.e., $\theta_{c}=0^{\circ}$.

Since truncation of forward scattering peak means considering the forward scattered photons as not interacting with the medium, the scattering opacity after truncation should be less than the scattering opacity without truncation as the forward scattering peak dominates the opacity. Note that $\kabs$ remains the same after truncation. As a reference, $\ksca$ for $\theta_{c}=0^{\circ}$, $1^{\circ}$, $2^{\circ}$, and $4^{\circ}$ are $\sim 4.4 \times 10^{-2}$, $3.6 \times 10^{-3}$, $3.3 \times 10^{-3}$ and $3.1 \times 10^{-3}$ cm$^{2}$ g$^{-1}$ respectively and $\kabs \sim 5.3 \times 10^{-4}$ cm$^{2}$ g$^{-1}$. The albedos are $\sim 0.99$, $0.87$, $0.86$, and $0.85$ respectively. 

To facilitate direct comparisons between the different models, we use $\Sigma_{c}=\tau_{0} / \kabs$ which fixes the absorption optical depth instead of the total optical depth as was done in previous sections. The benefit is that the total energy of emitted photons is kept the same and in the optically thin limit, the intensity should be the same regardless of $\ksca$. We set $\tau_{0}=0.1$ and use $10^{9}$ photons for all the models below. 

\begin{figure}
    \centering
    \includegraphics[width=\columnwidth]{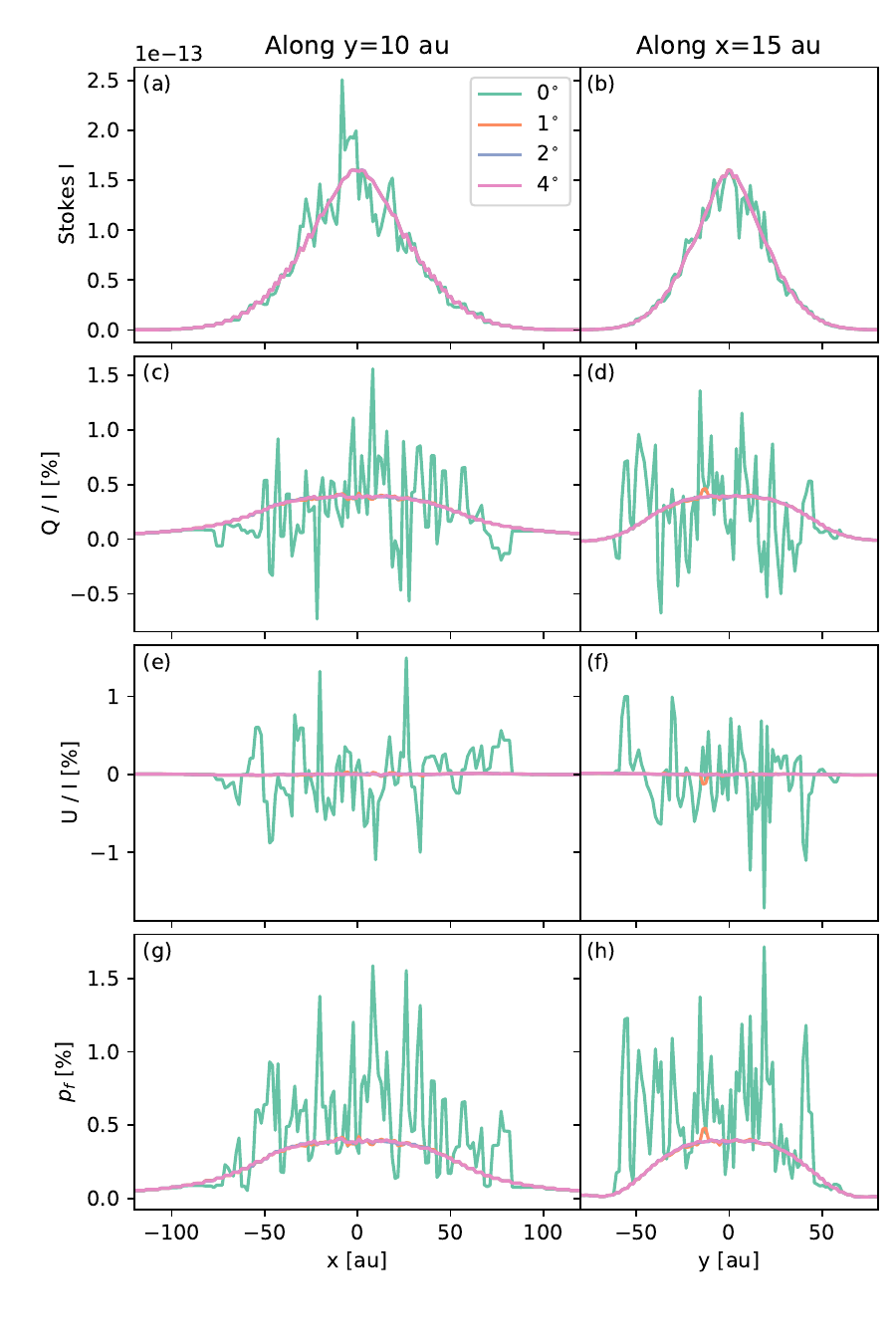}
    \caption{
        Comparisons of cuts along the images with different truncation angles $\theta_{c}$. The panels from top to bottom are Stokes~$I$ in erg s$^{-1}$ sr$^{-1}$ cm$^{-2}$ Hz$^{-1}$, $Q/I$, $U/I$, and $p_{f}$ in percent. The left column are cuts at constant $y=10$ au, while the right column are cuts at constant $x=5$ au. The profiles with $\theta_{c}=1^{\circ}$, $2^{\circ}$, and $4^{\circ}$ overlap each other given the quantitative similarities. 
    }
    \label{fig:chopdeg_cuts}
\end{figure}

In Fig.~\ref{fig:chopdeg_cuts}, we compare the models with different $\theta_{c}$ for two different cuts in the image. The first cut is along $y=10$ au, while the other is along $x=15$ au. For both cuts, the Stokes $I$, $Q$, $U$ and $p_{f}$ are smooth and quantitatively the same for $\theta_{c}=1^{\circ}$, $2^{\circ}$, and $4^{\circ}$. The model without truncation has not converged yet and the noise level for Stokes~$Q$ and $U$ are $\sim 1-2\%$ of Stokes~$I$. Nevertheless, Fig.~\ref{fig:chopdeg_cuts}a and b show that the Stokes~$I$ result without truncation is at the level of those with truncation. 

\begin{figure}
    \centering
    \includegraphics[width=\columnwidth]{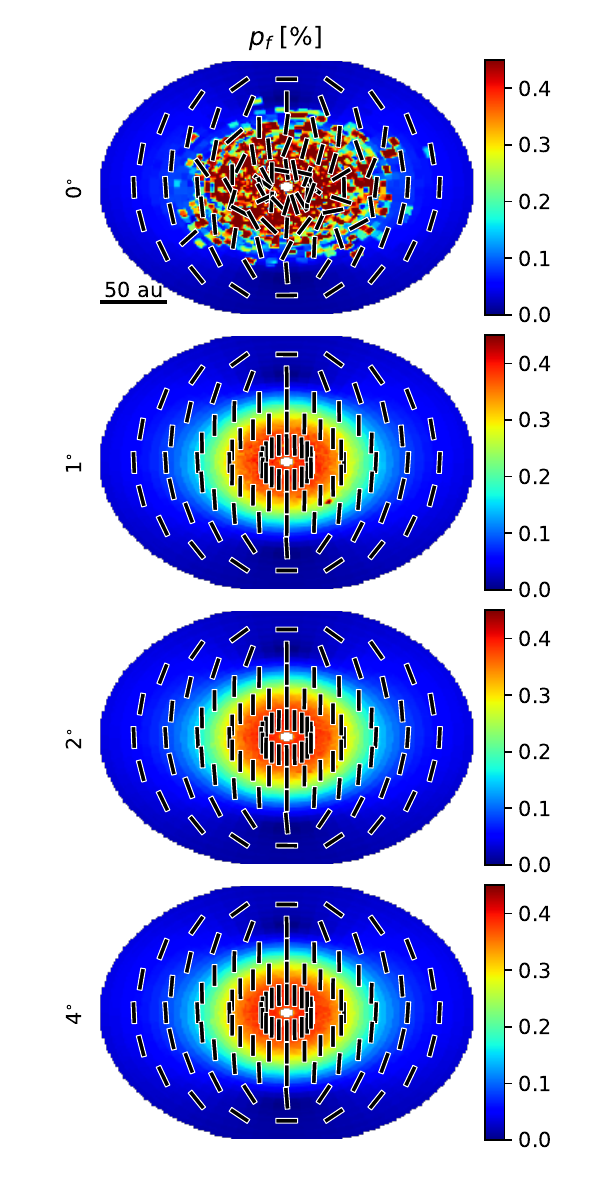}
    \caption{
        Comparisons of the $p_{f}$ images (color maps) and its polarization angles (line segments) for different truncation angles $\theta_{c}$ in $0^{\circ}$, $1^{\circ}$, $2^{\circ}$, and $4^{\circ}$ from top to bottom. 
    }
    \label{fig:chopdeg_im2d}
\end{figure}

Fig.~\ref{fig:chopdeg_im2d} shows the images of the polarization fraction and direction for each model. Fig.~\ref{fig:chopdeg_im2d}a clearly shows a highly noisy polarization image in contrast to Fig.~\ref{fig:chopdeg_im2d}b-d. The polarization directions in the center of Fig.~\ref{fig:chopdeg_im2d}a are also messy due to the noise. The noisy region is mainly in the center where the disk is more optically thick and requires more scattering to converge. 

The results are not too surprising, since the model without truncation scatters most of the photons in the forward direction which does not contribute much polarization, while the side-scattered photons that are responsible for providing polarization are rare. The necessity to truncate the forward scattering peak can be seen by comparing the total scattering opacity against the scattering opacity within the cone inside the truncation angle. The probability of photons scattered in $\theta \in [0, \theta_{c}]$ is
\begin{align}
    P(0 \leq \theta \leq \theta_{c}) = \dfrac{ 
            \int_{0}^{ \theta_{c} } F_{11}(\theta) \sin \theta d\theta 
        }{ 
            \int_{0}^{\pi} F_{11}(\theta) \sin \theta d \theta 
        }. 
\end{align}
As an example, the probability of photons that scatter within $1^{\circ}$ of forward scattering is $P(0^{\circ} \leq \theta \leq 1^{\circ}) \sim 95 \%$ for the XL sample. Thus, the vast majority of photons are forward scattered and only $\sim 5\%$ of the photons can contribute to polarization.


\bsp	
\label{lastpage}
\end{document}